\documentclass[onecolumn,aps,preprint,showpacs,amsmath,amssymb]{revtex4}
\usepackage{graphicx}
\usepackage{dcolumn}
\usepackage{bm}

\begin{document}
\newcommand{\eq}{\begin{equation}}                                                                         
\newcommand{\eqe}{\end{equation}}             

\title{Coherent control for the spherical 
symmetric box potential in short and intensive XUV laser fields} 

\author{ I. F. Barna$^1$ and  P. Dombi$^2$ }
\affiliation{$^1$ Atomic Energy Research Institute of the Hungarian Academy 
of Sciences, \\ (KFKI-AEKI), H-1525 Budapest, P.O. Box 49, Hungary\\
$^2$  Research Institute for Solid State Physics and Optics of the Hungarian 
Academy of Sciences, \\ (KFKI-SZFKI) H-1525 Budapest, P.O. Box 49, Hungary  \\}
\date{\today}

\begin{abstract}
Coherent control calculations are presented for a spherically symmetric                                    
box potential for non-resonant two photon transition probabilities.                                                            
With the help of a genetic algorithm (GA) the population of the excited 
states are maximized and minimized.                                                                              
The external driving field is a superposition of three intensive extreme 
ultraviolet (XUV) linearly polarized laser pulses with different frequencies 
in the femtosecond duration range.                                                
We solved the quantum mechanical problem within the dipole approximation. 
Our investigation clearly shows that the dynamics of the electron current has a                    
strong correlation with the optimized and neutralizing pulse shape.          
\end{abstract}
\maketitle

\section{Introduction}
Coherent control has become a routine procedure in physics and chemistry to optimize
and govern light-matter interaction processes in atomic and molecular systems \cite{wei,brix}.
However, widespread coherent control methods are limited to the shaping of visible or near-IR 
femtosecond laser pulses and to controlling corresponding transitions induced by them.
 There is, however, enormous future potential in this method. For example, extreme ultraviolet 
(XUV) pulses can be used to investigate hyper-fast inner shell processes in atoms and molecules 
and this possibility naturally raises the question whether a certain control can be also exercised 
by shaping the interacting attosecond waveforms. One of us has tried to answer this question by 
theoretically investigating a non-resonant two-photon transition in He (1s1s) - (1s3s) with shaped 
XUV pulses \cite{bar0}. 
Even though the answer was positive these calculations lacked insight into the fundamental physical 
background of the control process. 
Therefore, in this paper we investigate a more simple model system and analyze the control process for this 
parameter regime further. 

With the proliferation of tabletop, Ti:sapphire-laser-driven XUV sources based on high harmonic 
generation, experimental tools have also become available even to university groups to carry out 
similar coherent control like experiments. This can be achieved in two ways: i) either by phase shaping the XUV 
beam after the high harmonic generation process to achieve the desired temporal shaping effect 
\cite{stra} or ii) by exercising the well-known coherent control methods on the infrared laser 
pulse generating the harmonics. Even though, to our knowledge, this latter option has mainly been 
used to maximize harmonic conversion efficiency \cite{bart,yos,wal},
either by temporal or by spatial shaping of the genetic IR pulse. Option i) is also attractive 
since it provides a more direct control method and first pioneering experiments by Strasser {\it{et al}}. 
\cite{stra} have 
shown that this scheme is suitable for controlling coherent transients in a He atom. The 
demonstrated method also has the advantage of direct applicability to free-electron-laser beam-lines 
currently being developed. These novel light sources will also induce research along these lines in 
a broad spectral domain currently unavailable to laser-driven XUV/X-ray sources.

For the above-mentioned motivations we investigate the control process further with numerical tools 
for a more simple model system, a spherical box potential, and a simplified ansatz for the genetic 
algorithm. The motion of an electron was investigated in a spherically symmetric square well 
potential driven by a linearly polarized extreme XUV laser pulse. We solved the time-dependent 
Schr\"odinger equation with our simplified coupled-channel method which was successfully applied for 
more complex laser-atom interaction problems \cite{bar1}. Following the conventional experimental 
setups we apply the genetic algorithm (GA) \cite{car}. as optimization procedure to create the best 
interacting or most indifferent pulses (we call it neutralization) for state selective excitation. 
Our results show that the wave packet dynamic, the center-of-mass of the electron current is 
strongly correlated with the shape of the laser pulse, which gives a physical interpretation for 
the control mechanism, at least for this model potential problem. Section 2 shortly outline the 
theoretical background of our used model, followed by a compact description of the GA. Section 
3 present our results with an explanation. Atomic units [a.u.] are used through the paper
unless otherwise indicated.
	                                                                                                             
\section{Theory}       

We solve the general time-dependent problem with our simplified
coupled-channel approach to describe controlled laser driven excitation 
processes in the spherically symmetric box potential . The original method can 
be found in our former studies 
\cite{bar1},\cite{bar2}.
For the expansion coefficients of the time-dependent wavefunction the following  
differential equation system  holds  
\eq                                                                                                   
\frac{da_k(t)}{dt} = -i\sum\limits_{j=1}^{N} V_{kj}(t)e^{i(E_k-E_j)t}                                 
a_j(t)  \quad (k=1,...,N)                                                                             
\label{kopequ}                                                                                        
\eqe                                                                                                  
  where $E_k$ and $E_i$ are the eigenvalues of the box potential, and will be specified later.
  The   coupling matrix elements                                                                                                            
\eq V_{kj}(t) = \langle  \Phi_k |\hat{V}(t)| \Phi_j  \rangle.                                         
\label{copmatr}               
\eqe                  
are taken with the  well known eigenfunctions of the box potential.  
The probabilities for transitions into final excited states $j$ after the                                   
pulse are simply given by                                                                            
\eq                                                                                                   
P_j = |a_j(t = T)|^2                                                                    
\eqe            
where T is the duration of the pulse.                                                                                      
To get the total excitation probability the corresponding                                                  
channels $P_j$ must be summed up.  When a state selective excitation 
probability is controlled then the corresponding channel is considered only.
We have to mention, that the box potential has a continuous spectra as well, which 
can interpreted as an ionization spectra. In the following we concentrate on non-resonant two-photon 
excitation processes and neglect three-photon ionization yields which have negligible contributions  
in similar atomic systems \cite{bar0}.
We restrict ourselves to linearly polarized laser pulses parallel to the z-axis. 
The length gauge with the dipole approximation is applied                                              
\eq                                                                                                       
{V}(t) = - {\bf{E}}(t)\cdot {\bf{r}}.                                                 
\eqe                                                                                                      
                                                                                                           
To understand the control mechanism we took a simple model,                                       
and investigated the three-dimensional spherically symmetric square-well                                                    
potential. With the help of the width 'b' and the depth '-$V_0$' (which are the only
two parameters of this potential) the number of the bound states can be fixed.                             
We tune these parameters in such a way (b= 5 a.u., $V_0$= 5 a.u.) that only four bound states           
exist. The four state have different angular momenta from zero up to three.                                 
A detailed analysis of the problem can be found in any textbook \cite{schiff}.                                          
The wave functions inside the box potential are the well known spherical Bessel functions, 
and the energies can be found as solutions of different transcendental equations.                  
The four bound states have the following energies:                                                         
$E_{\ell=0}=-3.6 $ a.u.,$ \quad  E_{\ell=1}=-1.85 $ a.u.,$ \quad  E_{\ell=2}= -0.36 $ a.u.,$ \quad         
E_{\ell=3}=-0.05 $ a.u.                        

For external driving field strength we add three different frequencies and use a 
sin$^2$ envelope,                              
\begin{eqnarray}                                                                                          
 \vec{E}(t) = E_n \cdot \sin^2 \left(\frac{\pi t}{T} \right)  \left[                                       
 a_1\sin(\omega_1 t +  \delta_1) +  \right. \nonumber\\                                                                     
 \left.  a_2\sin(\omega_2 t + \delta_2 ) +                                                                        
 a_3\sin(\omega_3 t + \delta_3) \right]\vec{e}_z.                                                         
\end{eqnarray}                                                                                           
where the frequencies are fixed and the three amplitudes $a_{1,2,3}$ and                                  
phases $\delta_{1,2,3}$ are the free parameters optimized through  
the GA \cite{car}.                                                                         
 For $\omega_1 $ we took a quasi resonant two-photon frequency:                                             
$  (E_{\ell=2}- E_{\ell=0})/2 \approx \omega_1 = 1.52 $  a.u.                                            
The resonant frequency is 1.62 a.u. and when $\omega_1 $ is closer to the                                      
resonance then all the pulses excite the system with a large probability 
($P_2 > $  ten percent range ), and the electron dynamics between the 
optimized and neutralized cases have the same properties.
 On the other hand if $\omega_1$ is much farther from resonance then                                          
the optimization can not give us enough excitation, and the system 
remains practically in ground state.
For the two other frequencies we took $ \omega_2 = 1.40 \>$ a.u.  and                                   
$\omega_3 = 1.85 \> $ a.u.                                          
The GA freely varies the phases $|\delta_{1,2,3}| \le \pi $ and the amplitudes            
$|a_{1,2,3}| \le 2 $ which allows combinations                                                               
where the contribution of the quasi resonant frequency is suppressed letting the off-resonant                                  
frequencies to evolve their effect.                                                                        
Contrary to the widely used phase modulation, this pulse fabrication mechanism                         
does not conserve pulse energy, that is why all pulses have to be                                          
renormalized to a reference pulse                                                                          
\eq                                                                                                        
  \int_0^T  |\vec{E}(t)|^2 dt =  25.13  \> a.u.                                                               
\eqe                                                                                                       
in this pulse we took the quasi resonant frequency only,                                         
$ \omega = 1.52 $ a.u., with a large field strength $E_0 = 0.6 $ a.u., and  
$T = 400.53 $ a.u. pulse duration and $E_n$ is the normalization constant eq. (5).
This parameter set gives us massive excitation probabilities which                              
is needed for the further investigation.                                                                  

For time propagation we use a Runge-Kutta-Fehlberg method 
of fifth order embedding an automatic time step regulation. 
As an optimization procedure we used the celebrated genetic algorithm  
which works in the following way. GA represents each possible solution, or 
individual, with a string of bits, termed a chromosome. For example two possible
phases are represented as [01101101]. (For the sake of clarity we assume that each parameter
can only take $2^3 = 8 $ values in this example.)
The first generation of individuals is selected randomly. Typically, we use a 
population of size of 20 which is about a factor of three larger than the number of the optimisable 
variables. For each generation, the following steps are carried out: (i) all the individuals 
are evaluated and assigned a fitness value. In our case it means that we calculate the ionization or
state selective excitation probability resulting from each parameter configuration and call it
 fitness value. The next generation of individuals is chosen by applying three GA 
operators selection, mutation and crossover. (ii) The selection operator chooses which of the 
individuals from the present generation will be transfered to the next generation.
The individuals are ranked according to their fitness, and then selected randomly with a 
certain probability based on the fitness. A very fit individual thus receives a high probability 
and can be selected many times, while low-fitness individuals may not be selected at all. 
(iii) The mutation operator, which is used very seldomly selects a few individuals and replaces  
one  (randomly selected) bit in a chromosome randomly by 0 or 1 (e.g. creating the chromosomes
[11010100] from [11000000].) (iv) The crossover operator takes two individuals at a time and exchanges 
part of their chromosomes. For example two chromosomes [01101000] and [01001110] can create the 
chromosomes [01101110] and [01001000]. The use of the mutation and crossover operators ensures 
that the GA does not became in a local minimum or maximum. The fittest individuals, however 
always survives to the next generation what is called elitism. 
Steps (i)-(iv) are then repeated until no new solutions appear between two consecutive generations.
In the optimizations presented in this work, the GA typically converges after 40-70 generations.
     
\section{Results}                                                                      

The laser field has six free parameters that have to be optimized through the control process.                                                       
The GA needs further, technical parameters \cite{car}. 
We used 20 different pulses  per generation (population size)  and 
let the process run through 60 generation                         
to achieve convergence. This means that $20\times60 = 1200$ different                                       
pulses were checked to find the most favorable.                                                           
For the permutation probability ($\approx $ 1/population size )  = 0.05,                                   
for crossover probability = 0.4, and for creeping probability = 0.12 values were used
giving us good convergence and gain matching the commendation of the routine \cite{car}.                                         
We found the algorithm stable and robust against slight parameter variations.  

    
\begin{figure}
\vspace*{1.5mm}           
\scalebox{0.6}{\rotatebox{0}{\includegraphics{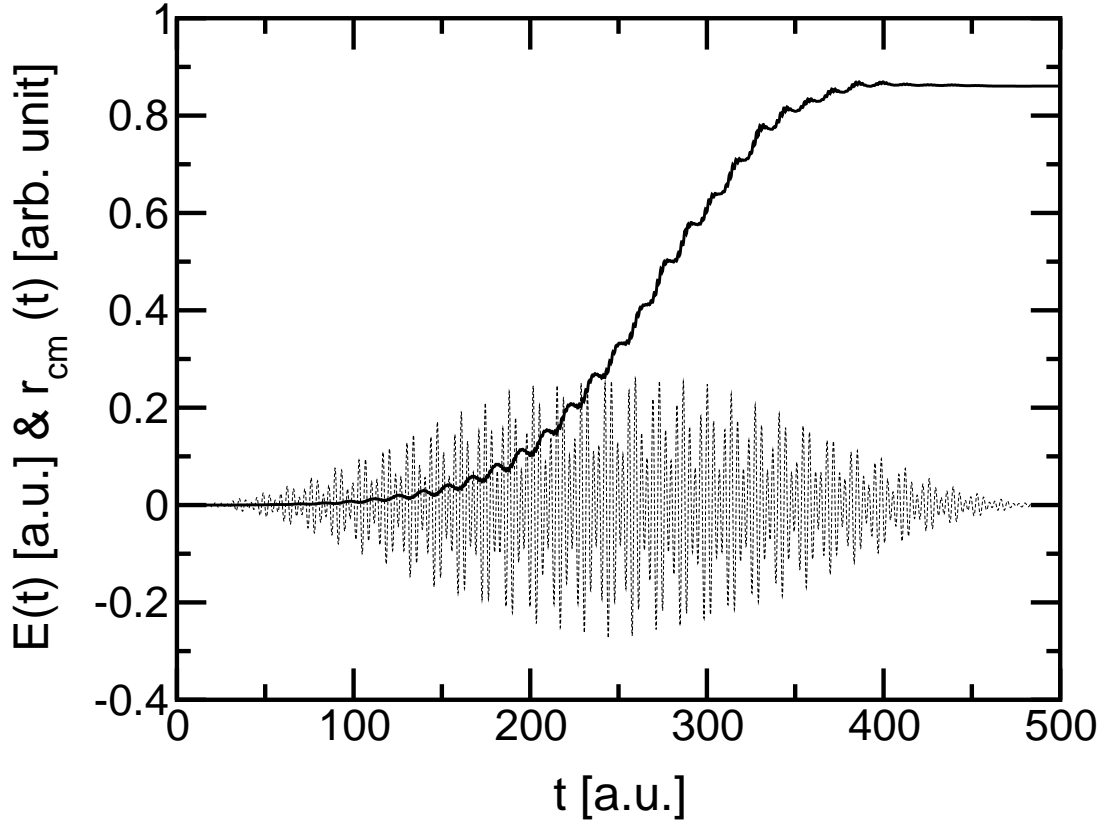}}}
\caption[]{\narrowtext
The electric field strength E(t) of the optimized pulse (dashed line)  with the corresponding
center of mass of the electron current $r_{cm}$(t)  solid line) }
\label{fig:gr}
\end{figure}


The GA makes it possible to carry out two different 
kinds of optimization calculations, first the maximization of the excitation
probabilities (we call it optimization) and secondly minimization (we call it 
neutralization) when the transition probabilities are minimized.
Both of these calculations had to be done, to interpret the 
control mechanism. 
Neutralization calculation may attract large interest in future FEL experiments                                      
where on one side the maximal field strength is used but on the other side                                
the field must not interact with the resonator causing any damage. 
The optimized $P_2$ transition probability for the pulse given above is $2.3 \times 10^{-3}$,
whereas for the neutralized probability it is $ 2.8 \times 10^{-5} $ which is a factor of 82 difference. 
The parameters of the optimized pulse are: $a_{1,2,3} \>\> [1.57, 0.68, 1.38]  \>\> 
\delta_{1,2,3} [0.45, -0.88, 0.52]$  
and for the neutralizing pulse: $a_{1,2,3} [1.17, 1.68, 0.82] $, $  \delta_{1,2,3}
 [-0.52, 0.66, 0.12]$   
                                                                                                           
To find some physical interpretation for our control results, we investigated                              
the dynamics of the electron wave packet.                                                                  
We calculated the time-dependent current of the electron through the following                             
formula:                                                                                                   
\eq                                                                                                        
{\bf{j}}({\bf{r}},t) =  \Psi({\bf{r}},t)^{\ast} \vec{\nabla} \Psi({\bf{r}},t)                       
\eqe                                                                                                       
for the time-dependent wave function we used the usual form:                                               
\eq                                                                                                        
\Psi({\bf{r}},t) = \sum_{\ell=0}^3 a_{\ell}(t) \Phi_{\ell}({\bf{r}})e^{-iE_{\ell}}                                               
\eqe                                                                                                       
where                                                                                                      
\eq                                                                                                        
\Phi_{\ell}({\bf{r}}) = \left( \frac{\pi}{2r} \right)^{\frac{1}{2}}\cdot                                            
J_{\ell+\frac{1}{2}}(rk) Y_{\ell,0}(\theta, \varphi).                                                               
\eqe               
The radial part $ \sqrt{\pi/2r} J_{\ell+0.5}(rk) $ 
 is the spherical Bessel function of the first kind, (usually noted with $j_{\ell}(rk)$ which may                                   
leads to miss-understanding in our case using the same letter for electron current)                                 
and $J_{\ell+\frac{1}{2}}(rk) $ is the ordinary Bessel functions of half-odd-integer order,                                         
and $ Y_{\ell,0}(\theta, \varphi) $ is the usual spherical harmonic.
                                                                               
The current is a complicated quick oscillating function, and it is difficult to                            
observe its correlation with the laser field, that is why                               
we introduce the center-of-mass coordinate of the electron current                               
\eq                                                                                                        
 r_{cm}(t) = \frac{\int j(r,t)\cdot r dr }{\int j(r,t) dr}                                                  
\eqe                                              
which is a pure time-dependent number. 
We concentrate on the radial current component only, skipping the vector notation.
It is easy to see, that the center-of-mass of the current is a bound function of time.                                
Before the laser pulse - without any external field - the system is in the $\ell =0 $                      
ground state with a wave function proportional to $\frac{\sin(rk)}{rk} $. This
function has maximum at zero and has a strong decay, hence
the center-of-mass coordinate is close to zero.                                                                  
On the other hand the wave functions of the excited states are higher order Bessel functions
having a zero at the origin and maximum at $r= 5.0 $ a.u.
Inside a laser pulse the electron is in a mixed quantum state.  
With a linear scaling we may identify the minimum of the current 
to zero and the maximum to a number comparable with field peak intensity.
Figure 1 shows the electric field strength of the optimized pulse
together with the corresponding electron current function. 
Beyond the sin$^2$ envelope shape and the carrier oscillation an extra modulation can be found 
with a quasi periodic time $\tau \approx 14 $ a.u. 
This property of the pulse is a beat phenomena.
The well-known addition theorem says:
\eq
 \sin(\alpha) + \sin(\beta) = 2\sin\left(\frac{\alpha+\beta}{2}\right)
 \cos\left(\frac{\alpha-\beta}{2}\right)
\eqe 
which means that adding two different frequencies with the same amplitude and the same phase 
gives us a new completely amplitude modulated signal with the beat time of:
$\tau_{beat} = \frac{\pi}{|\alpha-\beta|/2} $.
By checking the parameters of the optimized pulse, we found that $\omega_2$ and $\omega_3$  
have approximately the same amplitude and the same phase, giving us $\tau = 13.96 $ a.u.
The amplitude modulation is not maximal due to the existence of $\omega_1$.

Let's investigate the current now. It is clear to see that for $ t > 150 $ a.u. 
the current continously gains between two neighboring beat oscillation minima, and has a short 
plateau at the vicinity of the minima. 
This behavior can be explained as a resonance phenomenon, where the electron absorbs 
energy from the laser pulse at each intensity growth.
The time-dependent center-off-mass of the electron current shows two kinds of oscillations.
The slower oscillation follows the envelope of the pulse, (we call it envelope oscillation) 
and a much quicker but smaller oscillation follows the carrier of the pulse (we call it 
carrier oscillation). 
The work of Kosloff {\it{et al.}}\cite{kos} prove that there is a $\pi/2$ phase shift 
between the optimal pulse and the time dependent wave function overlap.
Figure 2 shows a magnified part of the field strength of the 
optimized pulse together with the overlap of the time-dependent ground sate wave function 
and the $\ell=2$ wave function through the dipole operator.
\eq
 O(t) = i\langle  \Psi_{\ell=2}({\bf{r}},t)|\hat{d} | \Psi_{\ell=0}({\bf{r}},t)    \rangle
\eqe
We optimized to a two photon transition that is why the overlap function has double
periodicity. The vertical line between the field strength and the overlap shows 
 $\pi/2$ phase shift. When the electric filed strength has a local minimum or  
maximum then 
the overlap function's carrier oscillation has a zero transition crossing the envelope
function. Due to the different time and amplitude scale of the envelope and carrier 
oscillation this effect is hard to see.

Finally, figure 3 presents the electric field strength of the neutralizing pulse
together with the corresponding electron current function. 
Contrary to the optimized pulse the current has local maximum when
the envelope reaches its minimum and vice versa.
The system decays with growing field strength.
This property of the current can be explained as an off-resonant dynamics,
field strength and electron movement.
The beat oscillation comes into play again $\tau_{beat} \approx 50/3 \approx 16.6 $ a.u.
which is hard to identify because $\tau_{\omega_1-\omega_3} = 19.03 $ a.u. and 
$\tau_{\omega_2-\omega_3} = 13.96 $ a.u.
No unambiguous correlation could be found between the second quick oscillation
of the current and the laser filed. 
It is important to mention that different optimized and neutralizing pulses 
and the corresponding currents were examined from the last generation
and all showed the same properties as analyzed above. 


\begin{figure}
\vspace*{1.5mm}           
\scalebox{0.6}{\rotatebox{-90}{\includegraphics{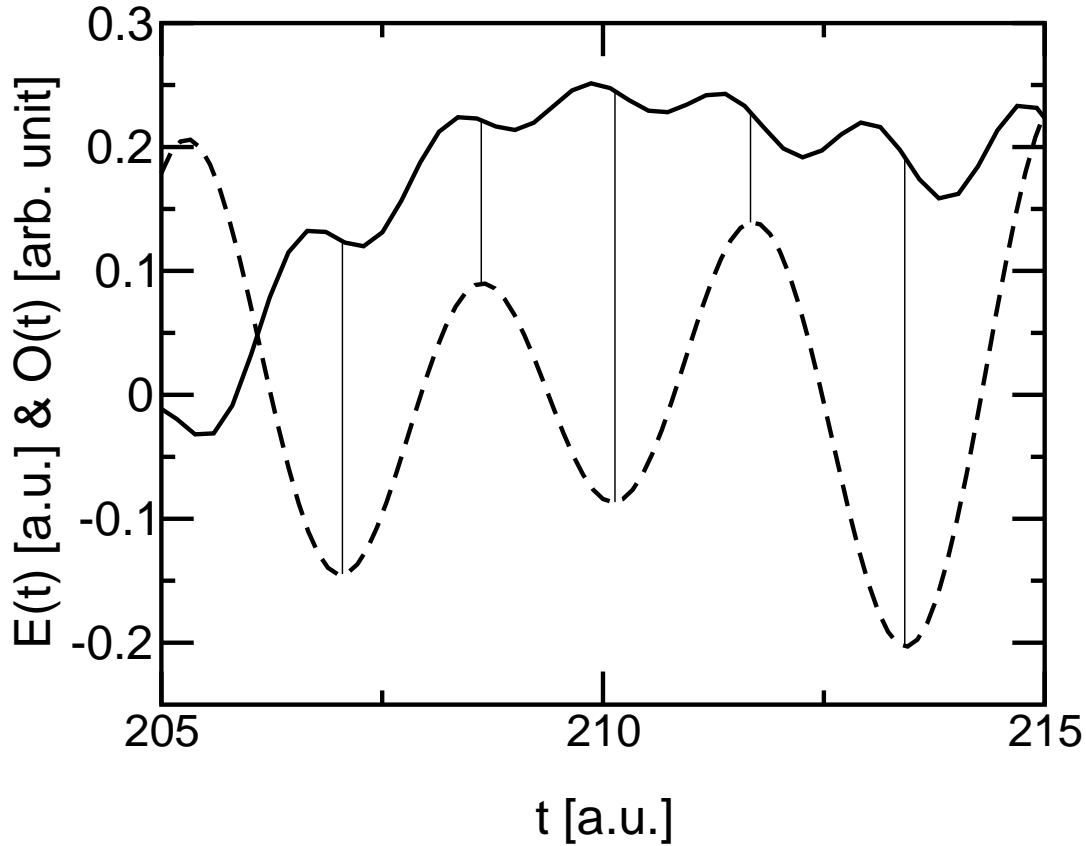}}}
\caption[]{\narrowtext
The $\pi/2 $ phase shift between the optimized time-dependent wave function
overlap O(t) (solid line)  and the electric field strength E(t) of the laser pulse  (dashed line)
\label{fig:ggg}}
\end{figure}


\begin{figure}
\vspace*{1.5mm}           
\scalebox{0.6}{\rotatebox{0}{\includegraphics{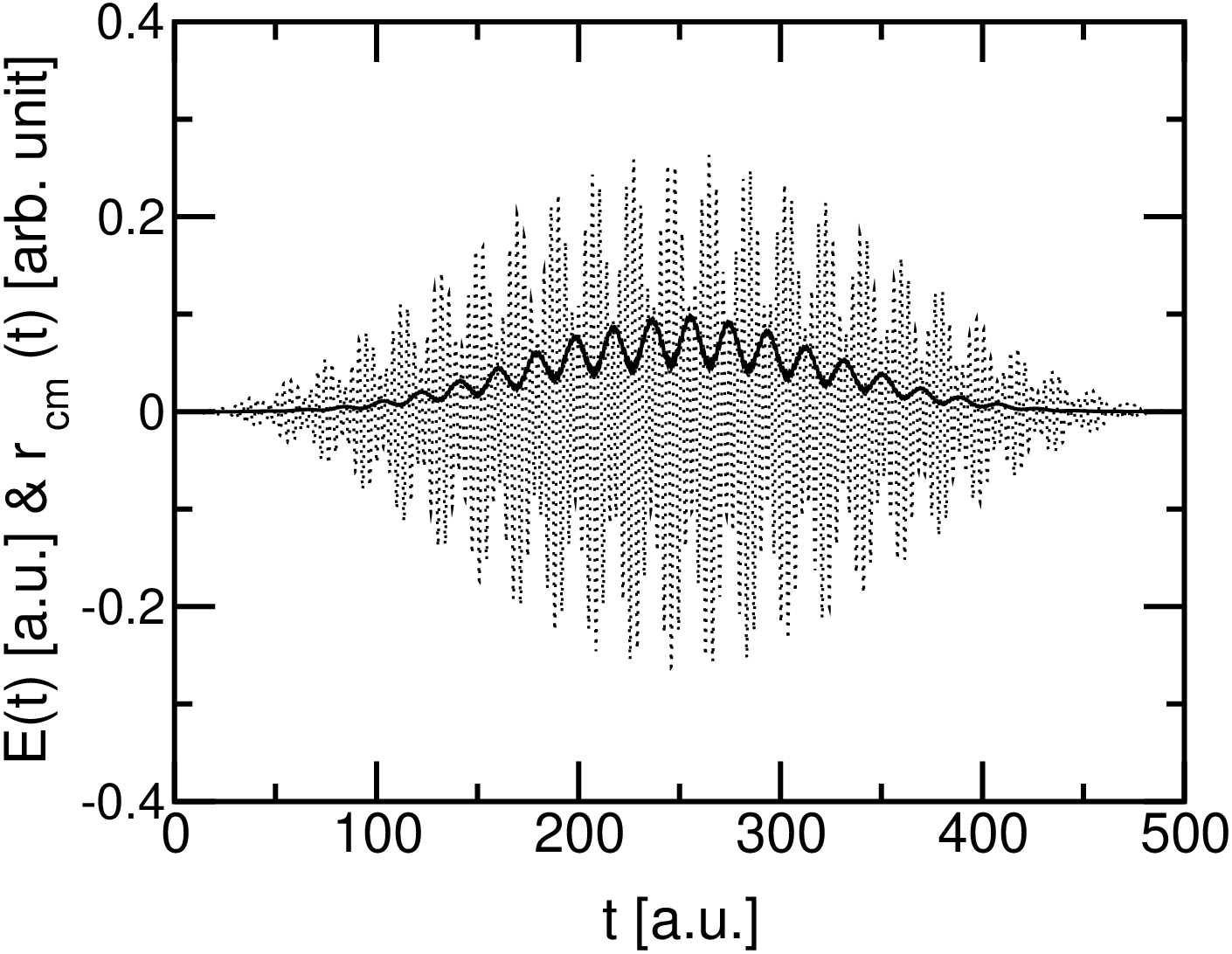}}}
\caption[]{\narrowtext
The electric field strength E(t) of the neutralized pulse (dashed line)  with the corresponding
center of mass of the electron current $r_{cm}$(t)  solid line) }
\label{fig:wo}
\end{figure}

                                                                                                           
\section{Summary and outlook}                                                                              
We presented coherent control calculations                                                  
for the spherically symmetric box potential in short intensive XUV laser pulses.
With the help of a GA we maximized and minimized two-photon non-resonant                              
probabilities. We found that the center-of-mass of the electron current is highly correlated with 
the envelope of the exciting laser pulse. The field of optimized laser pulses force the quiver 
motion on the electron wavepacket in-phase 
with the envelope of the pulse. However, and neutralized pulses force the electron to 
follow a motion which is out-of-phase relative to the envelope.                                                      

It is basically possible to implement our method for many-electron atoms too, to investigate 
and control ionization processes. Such works are in progress. Experimental coherent control experiments 
are carried out with visible or infrared lights nowadays. Our calculations were done in the hope 
that the rapid development of laser technology will make such schemes realizable in the near future.                             
Some results of first pioneering experiments are already available \cite{stra} with tabletop laser systems, 
moreover such control schemes are also applicable in free-electron-laser beam-lines delivering femtosecond 
XUV pulses.

\acknowledgements 
We thank Prof. A. Becker and Prof. J.M. Rost for fruitful discussions and constructive ideas. 
This work was supported by OTKA (Hungarian Scientific Research Found) F60256 and T048324. 
                                                                          
\vspace*{-5mm}



\begin{references}

\bibitem{wei} A.M. Weiner, Prog. Quant. Electr. {\bf{19}}, 161 (1995)                                       

\bibitem{brix}  T. Brixner, N.H. Darmrauer and G. Gerber                                                   
Advances in Atomic, Molecular, And Atomic Physics, Vol. {\bf{46}}, pp.                                     
1-46, (2001)                                                                                               


\bibitem{bar0} I.F. Barna  Eur. Phys. J. D                                                 
 {\bf{27}}, 287 (2003)                                                                                    

\bibitem{stra} D. Strasser {\it{ et al.}} Phys. Rev. A 
 {\bf{73}}, 021805(R) (2006)

\bibitem{bart} R. Bartels {\it{ et al.}} Nature  
 {\bf{406}}, 166 (2000)

\bibitem{yos} D. Yoshitomi {\it{ et al.}} Appl. Phys. B  
 {\bf{78}}, 275 (2004)

\bibitem{wal} D. Walter {\it{ et al.}} Opt. Express  
 {\bf{14}}, 3433 (2006)


\bibitem{bar1} I.F. Barna, J. Wang and J. Burgd\"orfer,  Phys. Rev. A                                                
 {\bf{73}}, 023402 (2006)                                                                                    

\bibitem{car} D.L. Carroll  Free Genetic Algorithm Driver                                                 
 {\it{http://cuaerospace.com/carroll/ga.html}}                                             

\bibitem{bar2} I.F. Barna {\it{Ionization of helium in relativistic                                       
heavy-ion collisions}}  Doctoral thesis, University Giessen (2002)                                      
{\it{``Giessener Elektronische Bibliothek"}}                                                            
 http://geb.uni-giessen.de/geb/volltexte/2003/1036                                                       

\bibitem{schiff} L.I. Schiff, {\it{Quantum Mechanics}} McGraw-Hill 1955                                  
Chapter IV, Page 76                                                                                       

\bibitem{kos} R. Kosloff {\it{ et al.}} Chem. Phys. {\bf{139}}, 201 (1989)                                                                        

\end{references}
\end{document}